\documentclass[preprint,amsmath,amssymb,nofootinbib]{revtex4}
\usepackage[dvips]{graphicx}
\usepackage{slashed}

\begin{document}

\title{A Lorentz-Violating Origin of Neutrino Mass?}

\date{April 12,2006}

\author{Andrew G. Cohen}
\email{cohen@bu.edu}
\author{Sheldon L. Glashow}
\email{slg@bu.edu}
\affiliation{Physics Department,  Boston University\\
  Boston, MA 02215, USA}

\begin{abstract}
  We explore implications for neutrino physics of Very Special
  Relativity (VSR), wherein the symmetry group of nature includes only
  a 4-parameter subgroup of the Lorentz group. VSR can provide a
  natural origin to lepton-number conserving neutrino masses without
  need for sterile (right-handed) states. Neutrinoless double beta
  decay is forbidden if VSR is solely responsible for neutrino masses.
  For ultra-relativistic neutrinos, such as are ordinarily studied,
  VSR and conventional neutrino masses are indistinguishable.  However,
  we show that VSR effects can be significant near the beta decay
  endpoint where neutrinos are not ultra-relativistic.
\end{abstract}

\maketitle

Very Special Relativity (VSR) is based on the hypothesis that the
space-time symmetry group of nature is smaller than the Poincar\'e
group, consisting of space-time translations and one of certain
subgroups of the Lorentz group\cite{Cohen:2006ky}. Here we focus on
implications of VSR for neutrino physics with the VSR subgroup chosen
to be the 4-parameter group SIM(2).

One might expect detectable consequences of a theory lacking Lorentz
invariance.  As noted in \cite{Cohen:2006ky}, most of the implications
of special relativity (Lorentz contraction, time dilation, isotropic
propagation of light in all inertial frames, {\it etc.\/}) are
preserved by SIM(2) invariance. Furthermore, the additional hypothesis
of CP invariance promotes SIM(2) invariance to full Lorentz symmetry.
Because the Lorentz invariant standard model works well, and because
violations of CP are small and have not been detected in the
flavor-diagonal sector, we expect VSR departures from Lorentz
invariance to be tiny. In this paper, we ask whether the relevant
parameter governing these effects may be related to the scale of
neutrino masses.

Consider the Lorentz invariant propagation of a free massive neutrino.
If its 4-component wave function $\nu$ satisfies the equation
$(\slashed{p}-m_{\nu}) \nu=0$, its Dirac mass is lepton-number
conserving, with $\nu_L\equiv \frac{1}{2}(1-\gamma_5)\nu$ a member of
a weak doublet and the additional state $\nu_R\equiv
\frac{1}{2}(1+\gamma_5)\nu$ a sterile singlet. Alternatively, if the
neutrino were to satisfy the equation $ \slashed{p}\,\nu_L -
m_{\nu}\,{\nu_L}^{c}=0$ (where the superscript $c$ denotes charge
conjugation) its Majorana mass would be lepton-number violating.

VSR admits the unconventional possibility of neutrino masses that
neither violate lepton number nor require additional sterile
states. As discussed in \cite{Cohen:2006ky}, the representations of
the little group of SIM(2) for massive states are one
dimensional. Consequently, VSR requires only two degrees of freedom
for a particle carrying lepton number: one for the neutrino with
lepton number 1, and one for the antineutrino with lepton number
$-1$. The VSR neutrino at rest is necessarily an eigenstate of angular
momentum in the preferred direction with eigenvalue $+1/2$. States
with any non-zero spatial momentum may be obtained from this state by
a VSR transformation.  The following equation captures these features:
\begin{equation}
  \label{eq:1}
  \left(\slashed{p} - \frac{m_{\nu}^{2}}{2}\frac{\slashed{n}}{p\cdot
    n}\right)\nu_L=0,     
\end{equation}
where $n$ is the light-like 4-vector $(1,0,0,1)$.  This equation is
manifestly not Lorentz invariant, but is invariant under SIM(2).  To
see this, recall that SIM(2), with its preferred direction chosen
along the $z$ axis, is generated by $K_x+J_y$ and $K_y-J_x$, along
with rotations and boosts along the $z$ direction ($J_z$ and $K_z$),
where $\mathbf{J}$ and $\mathbf{K}$ are rotations and boosts,
respectively.  The preferred 4-vector $n$ is changed in scale by
$K_z$, but is unchanged by the other generators of SIM(2). Thus the
term in \eqref{eq:1} proportional to $m_{\nu}^2$ and homogeneous in
$n$ is a SIM(2) invariant.

States in a massive unitary representation of SIM(2) (along with
space-time translations) are labeled by the invariant length of the
4-momentum, $m$, as in the Lorentz invariant case. This implies that
all massive particles, including the VSR neutrino, have a conventional
dispersion relation $p^{2}=m^{2}$. For the case at hand, this may be
verified by squaring the parenthesized operator in \eqref{eq:1} and
using the fact that $n\cdot n=0$.  Thus the neutrino acquires a
lepton-number conserving mass by construction, a result that cannot be
obtained in a Lorentz invariant context without introducing sterile
neutrinos. 

We have described a single neutrino state, but we may just as well
consider the physically relevant case of three active neutrino
species, whereupon the parameter $m_{\nu}^2$ becomes an arbitrary
non-negative Hermitean $3\times 3$ matrix.  Neutrino oscillation
phenomena remain entirely conventional in the VSR scenario: they are
described by the usual three mixing angles, one CP violating
parameter, two squared-mass differences (and two essentially
unobservable `Majorana phases'). While VSR can provide a novel origin
of neutrino mass, other mechanisms may contribute as well, such as a
see-saw involving heavy unobserved states.  However, guided by the
principle of simplicity, we adopt the tentative hypothesis that VSR is
the sole or dominant origin of neutrino mass.

Equation \eqref{eq:1} is not Lorentz invariant, implying (as we have
noted) that neutrinos at rest have spins pointing in the
preferred $z$ direction.  However, for ultra-relativistic neutrinos
($E/m \equiv\gamma \gg 1$), the consequent Lorentz-violating effects
are tiny ($\sim 1/\gamma^2$) except within a narrow cone about the $z$
axis with opening angle $\theta\sim 1/\gamma$.  Under practically all
realizable circumstances, no observable effects ensue.  {\it E.g.\/} a
VSR mass of 1~eV would modify the total cross-section of MeV neutrinos
off electrons by mere parts per trillion.  Fortunately, there is an
experimentally accessible venue in which neutrinos are not
ultra-relativistic: near the endpoint of the electron spectrum of beta
decay, where VSR neutrinos produce effects quite different from those
due to Lorentz invariant massive neutrinos.

For decades experiments have attempted to determine the electron
neutrino mass by means of precision studies of the electron spectrum
in beta decay near its endpoint.  The most sensitive experiments yet
performed provide only upper bounds to $m_{\nu}$, which at 95\%
confidence are
\begin{equation}
  \label{eq:3}
  m_{\nu_{e}} < 2.3\;\text{eV \cite{Kraus:2004zw}} \quad
    \text{and}\quad m_{\nu_{e}} <  2.5\; \text{eV
      \cite{Lobashev:2001uu}}\,. 
\end{equation}
The proposed KATRIN experiment\cite{Drexlin:2005zt} should be capable
of detecting and measuring a conventional (Lorentz invariant) electron
neutrino mass if it exceeds 0.2~eV.  In view of what has been learned
about neutrino masses from oscillation experiments, a positive result
to the KATRIN experiment would imply that the three neutrino species
are quasi-degenerate.  Thus neutrino mixing effects may be neglected
in the analysis of the data. Here we examine the consequences for this
experiment were neutrino masses to have a purely or dominantly VSR
origin.

For Lorentz invariant neutrinos, the effect of their mass on the
electron spectrum is due exclusively to the kinematic restriction of
phase space:
\begin{equation}
  \label{eq:4}
  \frac{dN/dE\lvert_{\text{conv}}}{dN/dE\lvert_{m_{\nu}=0}}
  =\frac{p_{\nu}}{E_{\nu}}
\end{equation}
where the neutrino energy $E_\nu$ and momentum $p_\nu$ are determined
by the electron kinetic energy $E$:
\begin{equation}
  \label{eq:5}
  E_\nu= E_{0} -E\quad\text{ and }\quad p_\nu=\sqrt{E_\nu^2-m_\nu^2}\,,
\end{equation}
with $E_{0}\simeq 18.574\  \text{keV}$,  the nominal endpoint, {\it
  i.e.\/} the maximum kinetic energy of the electron for a massless
neutrino.

For VSR neutrinos, however, the phase space restriction is augmented
by a change in the relevant matrix element.  From \eqref{eq:1} we see
that the weak leptonic charged current $j^{\mu}$ must be modified to
ensure its conservation:
\begin{equation}
  \label{eq:2}
  j^{\mu} = \bar{e}  \gamma^{\mu} \nu_{L} + \frac{m_{\nu}^{2}}{2} \bar{e}
  \frac{n^{\mu}\,\slashed{n}}{n\cdot p_{e}\;n\cdot p_{\nu}}\nu_{L}\ .
\end{equation}
The unconventional second term yields a correction of order
$E_{\nu}/m_{e}$, which is an entirely negligible effect near the
endpoint. More significantly, the first term, although not modified in
form, differs from the conventional current because the neutrino
spinor $\nu_{L}$ is different. The form for this spinor may be found
by solving \eqref{eq:1}.
The square of the matrix element for beta decay involves the VSR
neutrino spinor times its conjugate:
\begin{equation}
  \label{eq:7}
  \nu_{L}\bar\nu_{L} = P_{L}\left(\slashed{p} -
    \frac{m_{\nu}^{2}}{2}\frac{\slashed{n}}{p\cdot n} \right),
\end{equation}
in contrast to its conventional expression $P_{L}\slashed{p}$.
Averaging the square of the matrix element over electron spins,
multiplying by the phase space, and integrating over final state
angles, we obtain
\begin{equation}
  \label{eq:6}
  \frac{dN/dE\lvert_{\text{vsr}}}{dN/dE\lvert_{m_{\nu}=0}} =
  \frac{p_{\nu}}{E_{\nu}}\left(
    1-\frac{1}{4}\frac{m^{2}_{\nu}}{E_{\nu}p_{\nu}}\ln
    \frac{E_{\nu}+p_{\nu}}{E_{\nu}-p_{\nu}} \right)=  \frac{p_{\nu}}{E_{\nu}}\left(
    1-\frac{1}{2}\frac{m^{2}_{\nu}}{E_{\nu}p_{\nu}}\phi \right) ,
\end{equation}
where $\phi $ denotes the neutrino rapidity, $\cosh{\phi}=
E_\nu/m_\nu$.  The factor in parentheses---the VSR modification of
\eqref{eq:4}---increases monotonically from 1/2 at the endpoint toward
one.

\begin{figure}
  \centering
  \includegraphics[width=6in]{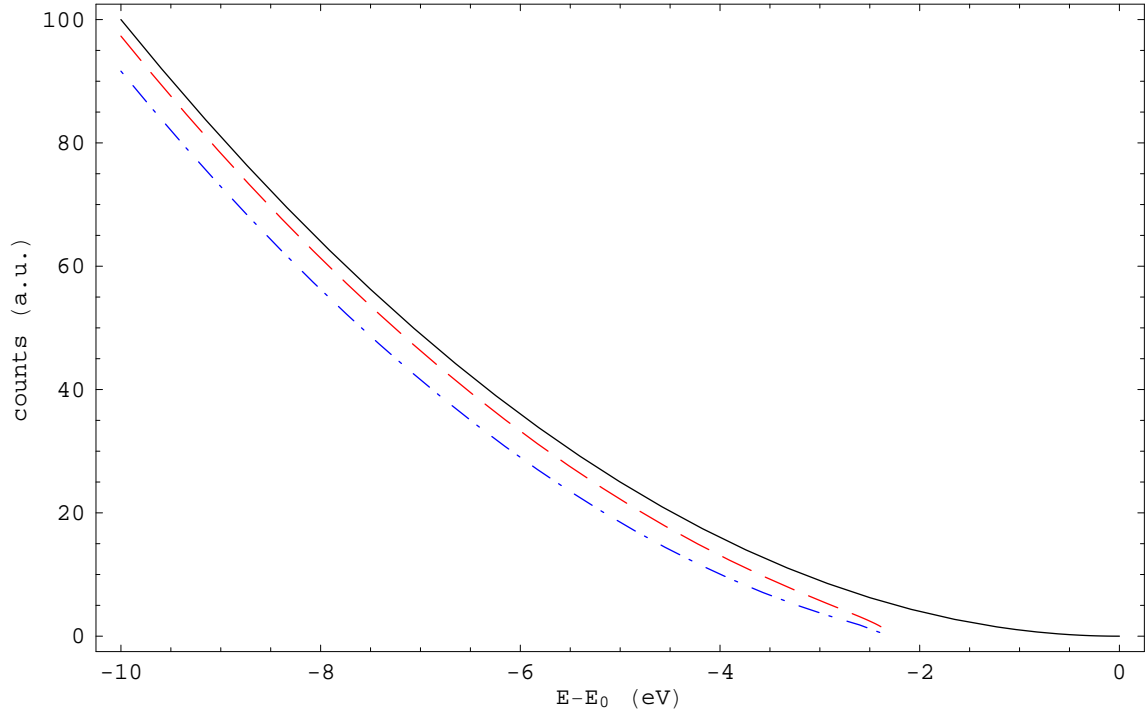}  
  \caption{$dN/dE$ near the endpoint of the tritium beta-decay
    spectrum.  The solid curve is for a massless neutrino; the dotted
    curve for one with conventional mass 2.3~eV; the dot-dash curve for
    one with the same VSR mass.}
  \label{fig:1}
\end{figure}

Our result for $dN/dE$ near the tritium endpoint is shown in Figure
\ref{fig:1}.  The upper curve describes a massless neutrino; the
middle curve a conventional 2.3~eV neutrino; and the lower curve a VSR
neutrino of the same mass.  Both the magnitude of the neutrino mass
effect and its energy dependence are different for the two cases. In
particular the present bounds on $m_{\nu}$ from endpoint experiments,
\eqref{eq:3}, are likely to be more stringent for VSR neutrinos.

\begin{figure}
  \centering
  \includegraphics[width=6in]{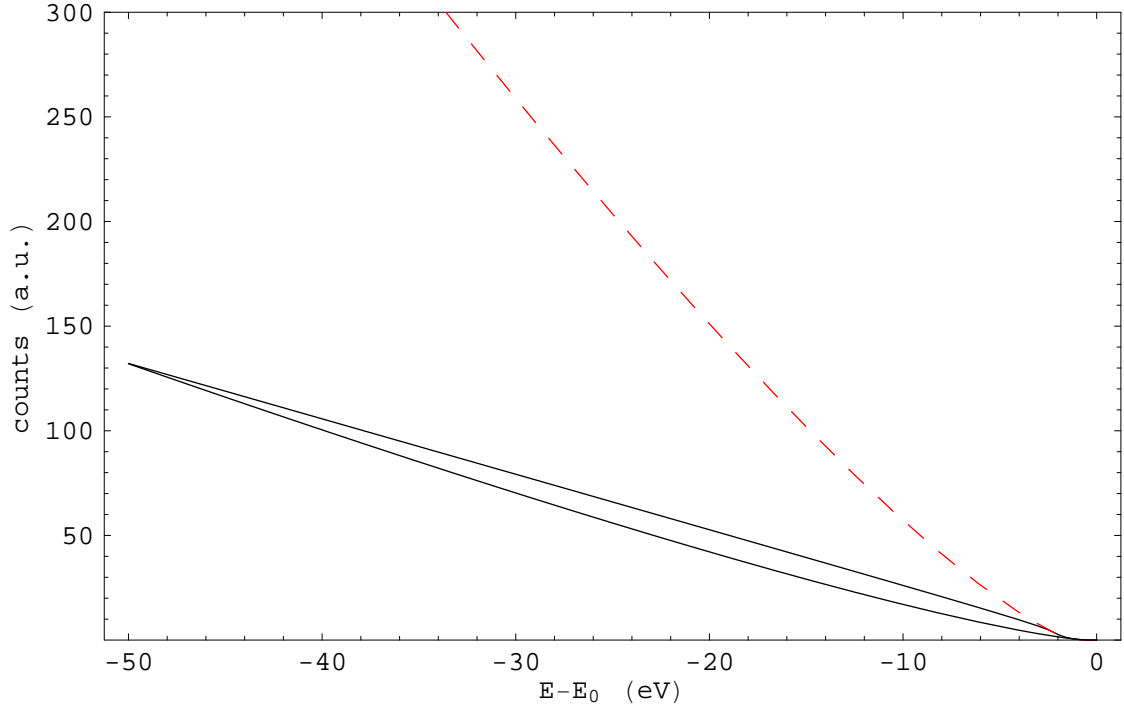}
  \caption{The integrated endpoint difference $R_{m_{\nu}=0}(E)-R(E)$
    for a neutrino with conventional mass 2.3~eV (lower curve), one
    with VSR mass of 2.3 eV (dashed curve), and one with VSR mass of
    1.08~eV (middle curve).}
  \label{fig:2}
\end{figure}

The KATRIN experiment and its predecessors measure the
\emph{integrated} energy spectrum from the endpoint downward:
\begin{equation}
  \label{eq:8}
  R(E) = \int_{E}^{E_{0}-m_{\nu}} \frac{dN}{dE'} dE' .
\end{equation}
The effect of neutrino mass on $R(E)$ is conveniently expressed as the
difference from the massless case.  Figure \ref{fig:2} shows this
difference for three cases. The central curve corresponds to a
conventional neutrino of mass 2.3~eV. The upper curve (truncated),
corresponding to a VSR neutrino of the same mass, shows that the VSR
effect on the difference is considerable.  For the lower curve, we
have adjusted the VSR mass to 1.08~eV so as to give an integrated
difference at 50~eV indistinguishable from that for a conventional 2.3
eV neutrino.

We see from Figure \ref{fig:2} that the effect on the electron
spectrum of a Lorentz invariant massive neutrino is similar to (but
not identical with) that of a VSR neutrino with about half the mass. A
future endpoint experiment that succeeds in detecting and measuring
the effects of neutrino mass may also be capable of distinguishing
between the conventional and VSR cases.

We have proposed an unconventional explanation for the origin of
neutrino mass in terms of Very Special Relativity. In contrast to the
usual Lorentz invariant scenario, VSR masses need not result from
Yukawa couplings. Furthermore no neutral states must be added, such as
the heavy neutrinos of see-saw models or the sterile partners of
neutrinos in the Dirac scheme. VSR neutrinos have two observable
implications for neutrino physics:
% We have shown that a VSR origin of neutrino mass has two immediate
% consequences for neutrino physics:
the absence of neutrinoless double beta decay, and a characteristic
modification of the tritium endpoint spectrum that may be detectable
by the next generation of endpoint experiments.  Weak $SU(2)$ symmetry
suggests that VSR may also have observable implications for charged
leptons.  The anticipated size of such effects for electrons,
$m_{\nu}^{2}/m_{e}^{2}$, suggests that they may be accessible to
sensitive atomic physics experiments\cite{prep}.  It would be truly
remarkable if experiments at very high precision and very low energy
were to provide the first indication of a failure of Lorentz
invariance.

\begin{acknowledgments}
  SLG thanks Yuval Grossman for an illuminating discussion.
  AGC was supported in part by the Department of Energy under grant
  no. DE-FG02-01ER-40676.
\end{acknowledgments}

\bibliography{vsr}

\begin{thebibliography}{5}
\expandafter\ifx\csname natexlab\endcsname\relax\def\natexlab#1{#1}\fi
\expandafter\ifx\csname bibnamefont\endcsname\relax
  \def\bibnamefont#1{#1}\fi
\expandafter\ifx\csname bibfnamefont\endcsname\relax
  \def\bibfnamefont#1{#1}\fi
\expandafter\ifx\csname citenamefont\endcsname\relax
  \def\citenamefont#1{#1}\fi
\expandafter\ifx\csname url\endcsname\relax
  \def\url#1{\texttt{#1}}\fi
\expandafter\ifx\csname urlprefix\endcsname\relax\def\urlprefix{URL }\fi
\providecommand{\bibinfo}[2]{#2}
\providecommand{\eprint}[2][]{\url{#2}}

\bibitem[{\citenamefont{Cohen and Glashow}(2006{\natexlab{a}})}]{Cohen:2006ky}
\bibinfo{author}{\bibfnamefont{A.~G.} \bibnamefont{Cohen}} \bibnamefont{and}
  \bibinfo{author}{\bibfnamefont{S.~L.} \bibnamefont{Glashow}}
  (\bibinfo{year}{2006}{\natexlab{a}}), \eprint{hep-ph/0601236}.

\bibitem[{\citenamefont{Kraus et~al.}(2005)}]{Kraus:2004zw}
\bibinfo{author}{\bibfnamefont{C.}~\bibnamefont{Kraus}} \bibnamefont{et~al.},
  \bibinfo{journal}{Eur. Phys. J.} \textbf{\bibinfo{volume}{C40}},
  \bibinfo{pages}{447} (\bibinfo{year}{2005}), \eprint{hep-ex/0412056}.

\bibitem[{\citenamefont{Lobashev et~al.}(2001)}]{Lobashev:2001uu}
\bibinfo{author}{\bibfnamefont{V.~M.} \bibnamefont{Lobashev}}
  \bibnamefont{et~al.}, \bibinfo{journal}{Nucl. Phys. Proc. Suppl.}
  \textbf{\bibinfo{volume}{91}}, \bibinfo{pages}{280} (\bibinfo{year}{2001}).

\bibitem[{\citenamefont{Drexlin}(2005)}]{Drexlin:2005zt}
\bibinfo{author}{\bibfnamefont{G.}~\bibnamefont{Drexlin}}
  (\bibinfo{collaboration}{KATRIN}), \bibinfo{journal}{Nucl. Phys. Proc.
  Suppl.} \textbf{\bibinfo{volume}{145}}, \bibinfo{pages}{263}
  (\bibinfo{year}{2005}).

\bibitem[{\citenamefont{Cohen and Glashow}(2006{\natexlab{b}})}]{prep}
\bibinfo{author}{\bibfnamefont{A.~G.} \bibnamefont{Cohen}} \bibnamefont{and}
  \bibinfo{author}{\bibfnamefont{S.~L.} \bibnamefont{Glashow}}
  (\bibinfo{year}{2006}{\natexlab{b}}), \eprint{in preparation}.

\end{thebibliography}

\end{document}